\documentstyle[prb,aps,multicol]{revtex}
\begin{document}

\draft
\date{July 2001}

\title{ 
Full counting statistics of a general quantum mechanical
variable}
\author{ Yu. V. Nazarov}
\address{Department of Applied Physics and Delft Institute of
Microelectronics and Submicrontechnology,\\ Delft University of
Technology, Lorentzweg 1, 2628 CJ Delft, The Netherlands}
\author {  M. Kindermann}
\address{Instituut-Lorentz, Universiteit Leiden,
P.\,O.~Box 9506, 2300 RA Leiden,
The Netherlands}
\maketitle

\begin{abstract}
We present here a quantum mechanical framework for
defining the statistics of measurements of $\int dt \hat A(t)$,
$A(t$) being a quantum mechanical variable.
This is a generalization of the so-called full 
counting statistics 
proposed earlier for DC electric currents. 

We develop an influence functional formalism that allows us 
to study the quantum system along with 
the measuring device thus fully accounting  
for the action of the detector on the system to be measured. 
We define the full counting statistics of an arbitrary
variable by means of an evolution operator
that relates initial and final density matrices of the measuring
device.

In this way we are able to resolve inconsistencies that 
occur in earlier definitions. 
We suggest two schemes 
whereby the so defined full statistics can be
observed experimentally.
\end{abstract}


\section{Introduction}
The measurement paradigm in quantum mechanics assumes
at least implicitly that the measurement is instant.\cite{Griffiths}
This is to be contrasted with a realistic measurement
of, say, electric current
where the result of measurement is averaged over
a sufficiently long time interval. If one intends to measure
a variable $A$, the individual measurement gives
$\int_0^T A(t) dt/T$. 
The reason for this is obvious: any measurement has to be accurate.
The integration over time reduces possible instant fluctuations
of $A(t)$ resulting in a more accurate outcome of an individual
measurement of this sort.
The dispersion of the probability distribution of the outcomes
is supposed to vanish in the limit of $T \rightarrow \infty$.
This paper focuses on the problems related to
the determination of this probability
distribution, the full statistics of the measurement results.

Several years ago Levitov and Lesovik \cite{Le1,Le2,Le3}
have made a significant step in the understanding
of this fundamental issue.
They have introduced a concept of full counting statistics (FCS)
of electric current and have found this statistics for
the generic case  of a  one-mode mesoscopic conductor.
The word "counting"  reflects the discreteness of
the electric charge. If electrons were classical particles,
one could just count electrons traversing a conductor.
The full counting
statistics could  be readily defined in terms of a probability
to have $N$ electrons transferred through the conductor
during a time interval $T$, $P_T(N)$. With this distribution
function one calculates the average current $\langle N \rangle /T$,
current noise
$(\langle N \rangle ^2> - \langle  N \rangle ^2)/T$ and all higher cumulants of the current.
A non-trivial value of interest is the probability
to have big deviations from the average value.
This can be measured with a threshold detector. The
probability distribution $P_T(N)$ would be the goal
of a quantummechanical calculation.

The operator of electric current through a conductor,
$\hat I$, is well-defined in the Fock space spanned
by the scattering states of electrons.
The initial idea of Lesovik and Levitov \cite{Le1} was to
define an operator of transferred charge by means
of a seemingly obvious relation
\begin{equation}
\hat Q_{tr} = \int_0^{T} dt \hat I(t).
\label{too_simple}
\end{equation}
Having this operator in hands, one applies
the general
paradigm of quantum measurement\cite{Griffiths}: The probability
to have a certain charge $q$ transferred
equals  the square of the 
projection of the wave function of the
system on the eigenstate of $\hat Q_{tr}$ with eigenvalue $q$.
Lesovik and Levitov were able to perform a challenging
calculation of these projections. However, they were hardly
satisfied with the results. For instance, the 
transferred charge was not quantized in units of the elementary charge.

This is why in the subsequent paper \cite{Le2}
the same authors
have proposed another method of evaluating $P_T(N)$.
Their scheme invoked an extra measuring device.
As a model device, they chose a precessing $1/2$ spin whose  precession angle should be  proportional to the transferred charge. The measurement paradigm is then
applied to the device. In this way they were able to
obtain a satisfactory definition of the full counting
statistics $P_T(N)$ with an integer number of
charges transferred. The details of the calculation
and a thorough discussion are presented in \cite{Le3}

It was clear to  the authors of \cite{Le3} that their
definition of the FCS does not
depend on a specific measurement scheme. However, this
fact was not explicitly evident. For several
years this hindered the impact of these outstanding
contributions.

One of the authors has recently proposed a slightly
different calculation scheme of FCS that does not implicitly
invoke any measuring device but still leads to the same results
\cite{Nazarov}. The observation
was that the cumulants of the current
can be obtained as non-linear responses of a system
to a fictitious field that can only be defined in the
framework of the Keldysh diagrammatic technique\cite{Keldysh}.
The calculation of FCS can be accomplished with a slight
extension of the Keldysh technique. This meant
some progress since the methods of the Keldysh technique are
well elaborated and can be readily applied to a variety
of physical systems and situations.

Recent work \cite{BelzigNazarov} has addressed the charge transfer
between two superconductors. The problem can be tackled with
an extension of the above-mentioned Keldysh technique. The expressions
for $P_T(N)$ were obtained. Albeit the authors have encountered
a significant difficulty with understanding the results in 
 classical terms. The point is that the calculation gave {\it negative}
probabilities. This indicates that the results cannot
be interpreted without invoking a quantum mechanical description
of a detector.

All this suggests that the quantum mechanical 
concept of full counting statistics shall be refined
and the generality of previously used definitions shall be
accessed. This is done in the present article.

To preserve  generality, we focus on the full counting
statistics of an arbitrary quantum mechanical variable
$A$. Then the result does not have to be discrete,
and, strictly speaking, no counting  takes place.
We keep the term "counting" for historical reasons.

We show that introducing a detector is absolutely
necessary for defining such statistics. On the other
hand, the answer does not depend on details of the
detector. This allows for a separation of the system measured
and the measuring device. We develop an exact quantum mechanical
description of the system in terms of a  path integral
over detector variables and derive our results from 
this description. We show that the classical interpretation
of full counting statistics is only possible in the presence 
of a certain symmetry. For the full counting statistics
of the electric current, this symmetry is gauge invariance.
In  superconductors the  gauge invariance is broken
and the full counting statistics must be interpreted along quantum mechanical
lines.

It is the main message of our paper that this interpretation
problem does not make the concept of full counting statistics
useless and/or unphysical. We show that it is the full counting
statistics that completely determines the evolution of the  density 
matrix of the detector. We show that thereby 
the statistics is observable in real experiments. We propose
and discuss two concrete measuring schemes.

The paper is organized as follows. 
We start with a general compact discussion of the interpretation
problems. We present the model of the detector 
in section \ref{model}. We present the approach and derive
 general relations for the detector  density
matrix in the subsequent section. Section \ref{Interpretation} defines the  FCS
and presents its interpretation. The subsequent sections provide
examples of FCS for an equilibrium system in the ground state,
a normal conductor and a harmonic oscillator.
The characterization of the FCS is considered in  sections 
\ref{sc:scheme1} and \ref{sc:scheme2} where two concrete measurement
schemes are discussed.
\section{General discussion}
\label{general_discussion}
It is not a priori clear why the operator definition 
(\ref{too_simple}) produces senseless results. We list below possible
intuitive reasons for this. To start with, the paradigm
 concerns an {\it instant } measurement.
The operator definition (\ref{too_simple}) 
is not local in time and  accumulates information about
the quantum state of the system for a (long) interval of time.
The applicability of the paradigm
is therefore not obvious. For instance, the averages 
of powers of $\hat Q_{tr}$ can be expressed in terms of correlators
of currents 
\begin{equation}
<\hat {Q}^N_{tr}> = \int_0^T dt_1 ... dt_N <\hat I(t_1) \hat I(t_2) ... \hat I(t_N)>
\label{Ncorrelator}
\end{equation} 
Usually the causality comes into quantum mechanics via
time ordering of operator products.
There is no  time ordering of current operators in (\ref{Ncorrelator}).
 This may indicate implicit problems with  causality.
The second reason is as follows. It seems obvious that the  time integral
of $\hat I$ can be associated with a physical operator of charge.
For an arbitrary operator $\hat A$ it may be difficult to find
such a physical associate. However, the integrals of $\hat A$ can be
measured, and statistics of those can be accumulated. 

The only way out seems to extend the original quantum mechanical 
system by adding a "detector". Although it is very physical,
mathematically this means an eventual doubling of the original Hilbert
space by adding extra degrees of freedom. In fact, this has
been done in \cite{Le2} by introducing the spin-$1/2$. Below we adopt
the simplest way to provide such an extension \cite{Emerged}.
Given a quantum mechanical system, one introduces an extra
variable $x$, whose operator $\hat x$ commutes with all operators
of the system. We assume that the canonically conjugated variable, $q$,
($[\hat x, \hat q ] = i$, in units with $\hbar = 1$)
can be measured according to  the paradigm. 
Next we introduce an interaction between the system and
the detector in such a way that in the time interval $(0,T)$
the Heisenberg equation of motion becomes
\begin{equation}
\dot{\hat  q}(t) = \hat A(t)
\label{correct}
\end{equation}
thus simulating Eq.\ (\ref{too_simple}).
In this way we avoid all possible difficulties with misinterpretations
of the paradigm. The integral of $A(t)$ is now correctly
associated with an operator of a value that can be {\it a priori} 
measured.

Albeit there is a price to pay. As we  show below, the FCS can be defined in this way as an operator that
relates the density matrices of the detector before and after the measurement.
In general, it is not the same as the probability distribution
of  the shifts of the detector momentum
which may be defined in the classical limit. The FCS can be interpreted in such terms
only under certain conditions, which are satisfied for the  statistics of
current in normal conductors. However, the definition we adopt
is sufficiently constructive since it allows to predict the  result
of any realistic measurement.

To make the detector more realistic and thus to show the generality 
of the results, one shall introduce internal dynamics of the detector
variable. This dynamics would make the detector a non-ideal one: 
the readings may differ from the definitions (\ref{too_simple}) and (\ref{correct}).
The path integral approach we describe below
provides the most convenient way to incorporate this internal dynamics.
\section{The Model}
\label{model}
Let us now specify our detection model. 
  The detector will formally be a  particle with the Hamiltonian
  $ \frac{\hat{q}^2}{2m} + V(\hat{x})$. $\hat{q}$ and $\hat{x}$ obey  canonical commutation relations.  
It shall be coupled to the system only for a certain period of time $T$
 and be decoupled adiabatically for earlier and later times.
 We introduce a smooth coupling function $\alpha_T(t)$ 
that  takes the value  $1$ in the time interval $0<t<T$ and 
is zero beyond the time interval $t_1<t<t_2$. The values for  $t_1<0$ and
$t_2>T$ are chosen in a way that  provides an  adiabatic switching.
  The whole Hamiltonian  therefore reads
\begin{equation}
 \hat{H}(t)= \hat{H}_{{\rm sys}}
 - \alpha_T(t)\hat{x}\hat{A} + \frac{\hat{q}^2}{2m} + V(\hat{x})
\end{equation}
We stress that this detector model is very general.
 It can  easily be  applied  to describe a variety of electrical circuits
  and mechanical systems. In particular, it may describe selective
  detectors: by choosing $V(x)=\frac{m}{2} \omega^2 x^2$
  we describe a harmonic oscillator that selectively detects
  a signal frequency $\omega$.
  In principle it would also be possible to treat detectors
   that consist of more than one degree of freedom in the  formalism
   we are going to develop.
 
The coupling is chosen  such  that the 
Heisenberg equation of motion for the detector quasimomentum
$\hat{q}$
\begin{equation} \label{eq:motion}
\dot{\hat{q}}(t)=\alpha_T(t) \hat{A}(t) - V'({\hat x}(t))
\end{equation}
reproduces Eq.\ (\ref{correct}) if the internal dynamics of
the detector is disregarded.
Eq.\ (\ref{eq:motion}) suggests that the statistics of measurements
of the  detector's  momentum
after uncoupling it from the system
will tell us about the full
counting statistics we are interested in.
The coupling term can be viewed as a disturbance of the system
measured by the measuring device.
To minimize this disturbance,
 we would clearly like to concentrate the detector wave function
 around $x=0$.  The uncertainty principle forbids us,
 however, to localize  it completely. Indeed, in such a way we would
  loose all information about the momentum,
 which we are going to measure.
  This is a fundamental limitation imposed by quantum mechanics,
  and we are going to explore
  its consequences step by step.
  To concentrate fully  on the
  quantum nature of the detector we  explore in details the limit of
  infinite detector mass. In this limit, $\dot{\hat{x}}=0$ and
  the detector can classically not disturb the system.  
\section{The Approach }
Our goal is to relate the density matrices of the detector
before and after the measurement. If there were no system to measure
we could easily express this relation in the form of a path
integral in the (double) variable $x(t)$, like it has been done many years
ago in \cite{Feynmann}. The expectation is that the
whole information about the system to be measured can be compressed
into an extra factor in this integral, the so-called influence functional.
 This functional will depend
on $x(t)$ rather than on the  internal dynamics of the detector. This
will make explicit the separation between the detector and the
measured system. To make contact with  \cite{Le3},
 we would like to present  this influence functional
as an operator expression that involves the system degrees of freedom 
only.
We denote the initial detector density matrix  (at $t<t_1$) by
$\rho^{in}(x^+,x^-)$ and the final one (at $t>t_2$, after having traced out the system's degrees of freedom) by
 $\rho^{f}(x^+,x^-)$. $\hat{R}$
 is the initial unperturbed
 density matrix of the system.
  The entire initial density matrix is assumed to factorize, $\hat{D}=\hat{R}\hat{\rho}^{in}$.
     
 We start out by inserting complete sets of states into
      the expression for the time development of the density matrix

\begin{eqnarray}
& \hat{\rho^{f}}(x^+,x^-) = \mathop{Tr}\limits_{\rm System} \nonumber \\
& <x^+|\overrightarrow{T}\exp\{-i \int_{t_1}^{t_2}{dt\;
\bigl[  \hat{H}_{\rm sys}-\alpha_T(t)\hat{x} \hat{A}+\frac{\hat{q}^2}{2m} +
V(\hat{x}) \bigr]}\}\;  \hat{D} \; \overleftarrow{T}\exp\{{i}
\int_{t_1}^{t_2}{dt\;
 \bigl[\hat{H}_{\rm sys}-\alpha_T(t)\hat{x} \hat{A}+\frac{\hat{q}^2}{2m}+
 V(\hat{x})\bigr] }\}|x^->
\end{eqnarray}
  in the usual manner.
 $\overrightarrow{T}(\overleftarrow{T})$ denotes  (inverse)
 time ordering. As the complete sets of states we choose  product
 states of  any complete set of states of the system and alternatingly complete sets of eigenstates of the position or the momentum operator of the detector.   Those intermediate states allow us to replace the position and momentum
 operators in the time development exponentials by their eigenvalues.
 We can then do the integrals over the system states as well as the momentum
 integrals and arrive at the expression
\begin{eqnarray}
 \hat{\rho^{f}}(x^+,x^-) = \int{ \mathop{D[x^+(t)]}\limits_{x^+(t_2)=x^+}\;\; \mathop{D[x^-(t)]}\limits_{x^-(t_2)=x^-} }\;\;\; \exp\{{i} \int_{t_1}^{t_2}{dt\; \frac{m}{2} \bigl[ (\dot{x}^+)^2-(\dot{x}^-)^2\bigr] + V(x^+(t)) - V(x^-(t)) \}}\;\;\mathop{Tr}\limits_{\rm System} \nonumber \\
 \overrightarrow{T}\exp\{-{i} \int_{t_1}^{t_2}{dt  \bigl[ \hat{H}_{\rm sys}-\alpha_T(t)x^+(t) \hat{A}\bigr] }\}\;\;  \hat{R} \;\; \overleftarrow{T}\exp\{{i} \int_{t_1}^{t_2}{dt \bigl[ \hat{H}_{\rm sys}-\alpha_T(t)x^-(t) \hat{A} \bigr] } \}  \rho^{in}(x^+(t_1),x^-(t_1))
\end{eqnarray}

 We rewrite the expression as 
\begin{equation}
\rho^{f}(x^+,x^-) = \int{dx_1^+ dx_1^- \;K(x^+,x^-;x_1^+,x_1^-,T) \rho^{in}(x_1^+,x_1^-)}
\end{equation}
in terms of the kernel 
\begin{eqnarray}
 K(x^+,x^-;x_1^+,x_1^-,T) = \int{ \mathop{D[x^+(t)]}\limits_{x^+(t_2)=x^+,x^+(t_1)=x_1^+} \;\;\mathop{D[x^-(t)]}\limits_{x^-(t_2)=x^-, x^-(t_1)=x_1^-}} \nonumber \\
\label{eq:kernel}
 F[x^+(t),x^-(t),T]\;\;  \exp\{{i} \int_{t_1}^{t_2}{dt\; \frac{m}{2} \bigl[ {\dot{x}^+(t)}^2-{\dot{x}^-(t)}^2\bigr] + V(x^+(t)) - V(x^-(t))}\}
\end{eqnarray}
that contains already  the desired influence functional
\begin{equation}
F[x^{+} (t),x^{-} (t),T] =  \mathop{Tr}\limits_{\rm System}
 \overrightarrow{T}\exp\{-{i} \int_{t_1}^{t_2}{dt \bigl[\hat{H}_{\rm sys}-\alpha_T(t)x^+(t) \hat{A}\bigr]}\}\;  \hat{R} \; \overleftarrow{T}\; \exp\{{i} \int_{t_1}^{t_2}{dt \bigl[ \hat{H}_{\rm sys}-\alpha_T(t)x^-(t) \hat{A}\bigr] }\} 
\label{time_dependent}
\end{equation}
We have therefore accomplished the task:  system and  detector
are presented as separate terms under the sign of path integration.

Seeking further simplifications,
we proceed by taking the limit of infinite detector mass
and considering a free particle, so that $V(x)=0$.
We find that the kinetic term in Eq.\ (\ref{eq:kernel})
then suppresses all fluctuations in the path integral.
The details of the calculation needed  to establish
this are provided in Appendix  A.
Therefore, the kernel $ K(x^+,x^-,x_1^+,x_1^-,T)$
becomes local in position space, does not depend on the  detector  and reads
\begin{equation}
 K(x^+,x^-,x_1^+,x_1^-,T)= \delta(x^+ - x_1^+)\; \delta(x^- - x_1^-) \;\; P(x^+,x^-,T) 
\end{equation}
with
\begin{equation}  \label{eq:local}
 P(x^+,x^-,T) =  \mathop{Tr}\limits_{\rm System}
  \overrightarrow{T} \exp\{-{i} \int_{t_1}^{t_2}{dt \bigl[ \hat{H}_{\rm sys}-\alpha_T(t)x^+
  \hat{A} \bigr] }\}
   \hat{R}  \overleftarrow{T}\exp\{\;{i} \int_{t_1}^{t_2}{dt \bigl[ \hat{H}_{\rm sys}-\alpha_T(t)x^-\hat{A} \bigr]}\} 
\end{equation}
If we compare expressions (\ref{time_dependent}) and (\ref{eq:local})
we see that the simplification is that the $x^{\pm}(t)$ do not depend on
time. So the detector is somehow "slow" and the variable $x(t)$  quasi-stationary.
This gives physical sense
to the limit we consider. The only requirement is eventually 
 time scale separation between the system and the detector.
The typical time scale of the internal dynamics of the detector
shall exceed that one of the system. Under these conditions, we can apply
the quasi-stationary approximation and replace $F$ with $P$.

It is constructive to rewrite now the
density matrices in Wigner representation
\begin{equation}
 \rho(x,q) = \int{\frac{dz}{2 \pi} \; e^{iqz}\; \rho(x+\frac{z}{2},x-\frac{z}{2})}
\end{equation}
and define
\begin{equation} \label{eq:P(x,q,T)}
 P(x,q,T)=  \int{ \frac{dz}{2 \pi} e^{iqz}\;P(x+\frac{z}{2},x-\frac{z}{2},T)}.
\end{equation}
This gives the following convenient relation
\begin{equation} \label{eq:dens}
 \rho^f(x,q) = \int{dq_1 \; P(x,q-q_1,T)
 \; \rho^{in} (x,q_1)}
\end{equation}
\section{Interpretation of FCS}
\label{Interpretation}
We adopt the relations (\ref{eq:local}),
(\ref{eq:P(x,q,T)}) and (\ref{eq:dens}) as the definition of
the FCS of the variable $ A$. Let us see why. First let us suppose
that we can treat the detector classically. Then the density
matrix of the detector in Wigner representation
can be interpreted as a classical probability distribution 
$\Pi(x,q)$ to be at a certain position $x$  with momentum $q$.
This allows for a classical interpretation of $P(x,q,T)$
as a probability to have measured  $q = \int_0^{T} A(t)$.
Indeed, one sees from (\ref{eq:P(x,q,T)}) that the final 
$\Pi(x,q)$ is obtained from the initial one by shifts in $q$,
$P(x,q,T)$ being the probability distribution of those shifts.

Albeit there is a point which hampers such classical interpretation.
The density matrix in Wigner representation
cannot be interpreted as a probability to have a certain position
and momentum or any other probability, since it is not positive.
Concrete calculations given below illustrate that  $P(x,q,T)$
does not have to be positive as well. Consequently, it cannot
be interpreted as a probability distribution. We stress that
the absence of a  classical interpretation does not hamper a "predictive
power" of FCS: having it in hands, one can predict the results
of measurements  using Eq.\ (\ref{eq:local}). The reverse problem
is to extract the FCS from the results of measurements. We discuss two schemes for this in sections \ref{sc:scheme1} and \ref{sc:scheme2}.

There is an important case where FCS can indeed
be interpreted as a probability
distribution. In this case $P(x,q,T)$ does not depend on $x$,
$ P(x,q,T) \equiv P(q,T)$. Then, integrating   Eq.\  (\ref{eq:dens})  over $x$, we find
\begin{equation}
\Pi^f(q)=\int{dq' \; P(q-q',T)\; \Pi^{in}(q')}
\end{equation}
with $ \Pi(q) \equiv \int{dx \rho(x,q)}$. 
Therefore, the FCS is in this special case the kernel that relates the probability distributions of the detector momentum before and after the measurement, $\Pi^{in}(q)$  and $\Pi^{f}(q)$, to each other. Those distributions are positive and so will  the FCS $P(q,T)$. Note, that a generic case in which this interpretation is appropriate is the measurement of a static operator $\hat{A}$, that is, $[\hat{H}_{\rm sys},\hat{A}]=0$. Then the FCS is obviously $x$-independent and it can be interpreted as the probability distribution of outcomes of an instant measurement of $\hat{A}$. 

 
Since we study FCS of a stationary system and the measurement
time $T$ exceeds timescales associated with the system,
the operator expression in Eq.\ (\ref{eq:local}) can be seen
as a product of terms corresponding to time intervals.
Therefore  in this limit of $T \rightarrow \infty$
the dependence on the  measuring time
can be reconciled into
\begin{equation}
P(x^+,x^-)= e^{-{\cal E}(x^+,x^-) T}
\end{equation}
where the expression in the exponent is supposed to be big.
Then the  integral (\ref{eq:P(x,q,T)}) that defines FCS can be done in
the saddle point approximation. If one defines $A=q/T$,
that is,  $A$ is the result of measurement of $\int A(t) dt/T$,
the FCS can be recast into the similar from
\begin{equation}
P(x,A,T) = e^{-{\tilde{\cal E}}(x,A) T}
\end{equation}
where ${\tilde{\cal E}}$ is defined as a (complex)
extremum with respect to (complex) $z$:
\begin{equation}
{\tilde{\cal E}}=\mathop{{\rm extr}}_{z} \{ {\cal E}(x+\frac{z}{2},x-\frac{z}{2}) - i A z\}
\end{equation}
The average value of $A$ and its variance (noise) can be expressed
in terms of derivatives of ${\cal{E}}$:
\begin{equation}
<A> = \lim_{z \rightarrow 0} \frac{\partial {\cal E}
(x+z/2,x-z/2)}{i \partial{z}}; \ \ T \ll A^2 \gg = \lim_{z
\rightarrow 0} \frac{\partial^2 {\cal E}
(x+z/2,x-z/2)}{\partial{z^2}}. 
\end{equation}
Indeed, the quantity $P(x^+,x^-,T)$ is  the generating
function of moments of $q$. It is interesting to note
that in general this function may generate a variety
of moments that differ in the time order of operators involved,
for instance,
\begin{equation}
Q^N_M = (-i)^N \lim_{x^\pm \rightarrow 0}
\frac{\partial^M} {\partial (x^-)^M}
\frac{\partial^{N-M}} {\partial (x^+)^{N-M}} P(x^+,x^-)=
\int_0^T dt_1 ... dt_N
<\overleftarrow{T} \{ \hat{A}(t_1)...\hat{A}(t_M)\}
\overrightarrow{T} \{\hat{A}(t_{M+1})...\hat{A}(t_N)\}>
\end{equation}
The moments of $q$ are expressed through these moments
and binomial coefficients
\begin{equation}
Q^{(N)} \equiv \int dq q^N P(0,q,T) = 2^{-N} \sum_M C^N_M Q^N_M
\end{equation}
\section{FCS of an equilibrium system in the ground state}
 To acquire a better understanding of the general relations
 obtained we consider now an important  special case.
 We will assume that the system considered is in its
 ground state $|g>$, so that its initial density
 matrix is $\hat{R}=|g><g|$. This assumption allows for
 an easy calculation of the FCS.
  We have supposed that the coupling between the system
  and the detector arises adiabatically.
  Then the time
  development operators in (\ref{eq:local}) during the 
  time interval $t_1<t<0$ adiabatically transfer
  the system  from $|g>$ into the ground state $
  |g(x^{\pm})>$ of the new Hamiltonian
  $\hat{H}_{\rm sys}-x^{\pm}\hat{A}$ .
  In the time interval $0 < t < T$
  the time evolution of the resulting state has thus the
  simple form 
\begin{equation}
 \exp \{ - {i}\;t\;(\hat{H}_{\rm sys}+x^{\pm}\hat{A})\}\;
|g(x^{\pm})>\;=\; e^{-{i}tE(x^{\pm})}\;|g(x^{\pm})>.
\end{equation}
 $E(x^{\pm})$ are the energies corresponding
 to $|g(x^{\pm})>$.
 This provides the main contribution to the
 FCS if the measurement time is large and the phase acquired
 during the switching of the interaction
 can be neglected in comparison with this contribution.
 Therefore we obtain
\begin{equation} \label{eq:ground}
  P(x^+,x^-,T) =  e^{- {i} T (E(x^+)-E(x^-))}.  
\end{equation}
First we observe that $P(x,q,T)$ is a real function
in this case, since the exponent in (\ref{eq:Pground})
is anti-symmetric in $z$.
A first requirement for being able to interpret $P(0,q,T)$
as a probability distribution is therefore fulfilled.
However, the same asymmetry assures that all {\it even}
cumulative moments of $q$ are identically zero, whereas
the odd ones need not. On one hand, since the second
moment corresponds to the noise and the ground state cannot
provide any, this makes sense. On the other hand,
this situation would be impossible if $P(0,q,T)$
were a true, positive probability distribution.

Belzig and Nazarov \cite{BelzigNazarov} encountered this situation
analysing the FCS of a general super-conducting junction.
In a certain limit the junction becomes a simple
Josephson junction in its ground state. In this mere limit
the interpretation of FCS as a probability distribution
does not work any longer.
Fortunately enough, the relation
(\ref{eq:dens}) allows us to interpret
the results obtained.

To keep simple things simple, we assume the function $E(x)$ to be analytic
and expand it in its Taylor series. We also rescale $q$ as above, $q=A/T$. For the FCS
we therefore have
\begin{equation}  \label{eq:Pground}
 P(x,A,T)=  \int{\frac{dz}{T}\; e^{iz A T}\;\cdot\; e^{-{i} T( E'(x) z + E'''(x) z^3/24 + ...)}}
\end{equation}
If we take the limit $T \rightarrow \infty$ now, we note that all
the terms involving higher derivatives of $E(x)$ are negligible and arrive at
\begin{equation} \label{eq:Pinfinite}
 \lim_{T\to\infty} \;P(x,A,T)\;= \delta(A - E'(x))
\end{equation}
According to the Hellman-Feynman theorem $E'(x)=<g(x)|\hat{A}|g(x)>$.
As one would expect, in this limit the measurement gives the expectation value
of the operator $\hat{A}$ in a ground state of the system
that is somewhat altered by its interaction  with the detector
at  position $x$. Therefore the  resulting dispersion of $A$ will be determined
by the {\it initial} quantum mechanical spread of the detector wave function.
Consequently, the error of the measurement comes from the interaction with
the detector rather than from the intrinsic noise of the system measured.
\section{FCS of the electric current in a normal conductor}
A complementary example is a normal conductor biased at finite
voltage. This is a stationary { \it non-equilibrium} system far from being
in its ground state. Here we do not intend to go to microscopic details
of the derivation. Our immediate aim is to make contact with the approaches of
Refs. \cite{Le3,Nazarov}. We keep the  original notations of the references wherever
it is possible.

The starting points of the approaches differ much. Levitov and Lesovik
propose a detector model where the $z$-component of a spin-$1/2$ creates a local
vector potential which is felt by the electrons. This corresponds to a total
Hamiltonian of the form
$$
\hat{H} = \hat{H}_{\rm sys} - \frac{\lambda}{2e} \hat{\sigma}_z \hat{I}
$$
which is studied at different coupling constants $\lambda$. The reference
\cite{Nazarov} starts with an extension of the Keldysh technique to formally
unphysical systems where the evolution of the wave function in different
time directions is governed by two different Hamiltonians
\begin{equation}
\hat{H}^{\pm} = \hat{H}_{\rm sys} \pm \chi \hat{I}
\end{equation}
and shows that the so defined Green functions can be used to generate moments
of $\hat{I}$. This is to be compared with our detection model.

Despite different starting points,
 all three approaches quickly concentrate on the  calculation
of the quantity
\begin{equation}
<\exp(-i \hat{H}_1 T) \exp(i \hat{H}_2 T)>
\end{equation}
which is identical to our definition of $P(x^+,x^-)$ provided that
$H_{1,2} = \hat{H}_{\rm sys} -x^{\pm} \hat{I}$.
This quantity is denoted by $\chi(\lambda)$ in \cite{Le3} and by $\exp(-S(\chi))$
in \cite{Nazarov}. It is used to define the FCS and
we see now that the final result does not depend on the starting point.

%

To give a concrete expression for FCS, we just make use of
Eq.\ (37) of \cite{Le3}. We consider the FCS of the current
in a multi-mode constriction which is characterized by a set
of transmission coefficients $T_n$. In general, the answer 
is expressed in terms of energy-dependent electron filling factors
$n_{R(L)}$ on the right (left) side of the constriction
\begin{equation}
P(x^+,x^-,T) = \exp \{ \frac{T}{2\pi} \sum_n \int^{+\infty}_{-\infty} dE \ln
(1+T_n (e^{i(x^{-}-x^{+})/e}-1) n_R (1-n_L) +T_n (e^{i(x^{+}-x^{-})/e}-1) n_L (1-n_R))\}
\end{equation}

We stress that this expression depends on $x^+ - x^-$ only. This is  a
direct consequence of gauge invariance. Indeed, in each of
the Hamiltonians the coupling term is the coupling to a vector potential
localized in a certain cross-section of the constriction. The gauge
transform that shifts the phase of the wave functions by $x^{\pm}/e$
on the right side of the constriction, eliminates this coupling term.
This transform was explicitly implemented in \cite{Nazarov}.
Since there are {\it two} Hamiltonians in the expression, the coupling
terms cannot be eliminated simultaneously provided $x^{+} \ne x^{-}$.
However, the gauge transform with the phase shift $(x^{+} + x^{-})/2$
makes the coupling terms depending on $x^+ - x^-$ only.

Since $P(x^+,x^-,T)$ depends on $x^+ - x^-$ only,
the FCS $P(x,q,T)$ does not depend on $x$. As we have stressed
in section \ref{Interpretation}, this enables one to interpret the FCS
as a probability distribution.

Superconductivity breaks gauge invariance, thus making such an interpretation
impossible.
\section{FCS of a harmonic oscillator} 
Let us now  illustrate the measuring process proposed  with a simple example. We consider  the  measurement of the position  of a harmonic oscillator  in its ground state. That is, $\hat{H}_0= \frac{\hat{Q}^2}{2M}+\frac{1}{2} M \omega^2 \hat{X}^2 $ is the Hamiltonian of the system and   $\hat{A}=\hat{X}$ shall be measured. The measurement is effected by coupling the oscillator to a detector with position variable $x$ during a time interval of length $T$.  In the limit of infinite detector mass the entire Hamiltonian in the measurement period reads then
\begin{equation} \label{eq:Ham}
 \hat{H}= \frac{\hat{Q}^2}{2M}+\frac{1}{2} M \omega^2 \hat{X}^2 +  \hat{x} \;\hat{X}.
\end{equation}
 The perturbed ground state $|g(x)>$ is in this simple example obtained by   shifting the original ground state wave function by $x/M\omega^2$  in $X$-representation. Its energy is $E_g(x)=E_g(0) - \frac{1}{2M\omega^2}\; x^2$ . We then find from (\ref{eq:Pground}), that 
\begin{equation} \label{eq:oscFCS}
 P(x,q,T) =  \delta(q+ x T/M\omega^2  )
\end{equation}
Following our first classical interpretation of $P(0,q,T)$ we would now conclude, that a harmonic oscillator in its ground state does not transmit any fluctuations of its position variable to the detector and that the detector's wave function is not  altered by the oscillator. Let us, however, calculate the read off of the detector after the measurement just described, i.e., its final momentum distribution. We have to integrate the final density matrix  found from  (\ref{eq:dens}) over $x$. As the initial detector state we choose a Gaussian wave with uncertainty $\triangle q $ in the momentum:
\begin{equation}
\rho^{in} ( x,q) = \exp \bigl[-\frac{q^2}{2(\triangle q)^2}- 2 (\triangle q)^2 x^2 \bigr] 
\end{equation}
The final momentum distribution
\begin{equation}
 \Pi^f(q) = \exp \bigl[ \frac{- q^2}{2(\triangle q)^2+ T^2/ 2 M^2 \omega^4(\triangle q)^2} \bigr]
\end{equation}
is again a Gaussian. Its width $\triangle q^f$ , however, increases in time, for big times linearly with $T$:
\begin{equation}  \label{eq:delp}
(\triangle q^f)^2 = (\triangle q)^2 +  \frac{ T^2}{4 M^2 \omega^4(\triangle q)^2}
\end{equation}
This relation clearly demonstrates the relevance of the detector's influence on the system due to its quantum nature.
The initial detector momentum is known only with an uncertainty $\triangle q$. This uncertainty is a source of measurement error and is described by the first term of Eq.\ (\ref{eq:delp}). We would clearly like to minimize this error by choosing  $\triangle q$ to be small. Doing so, we increase, however, the uncertainty in the detector position $\triangle x \gtrsim 1/2 \triangle q$. According to Eq.\ (\ref{eq:Ham}), such a spread in the detector position will disturb the system. Since the oscillator is in its ground state this disturbance will drive it into  ground states of  new Hamiltonians $\hat{H}_0 + x \hat{X}$.  For every detector influence  $x$ we  will therefore measure a different expectation value $ E'(x)=<g(x)|\hat{X}|g(x)>$. The read off of the detector will  be a superposition of  measurement outcomes corresponding to all those different influences. The resulting broadening of the final detector momentum distribution  $ (\triangle q^f)_{\triangle x} \approx \triangle x \; E''(0) T$ is  accounted  for by the second term in (\ref{eq:delp}).  We conclude that the the quantum fluctuations of the  detector set an upper bound on the accuracy of the measurement process. The actual value of this  bound depends of course on the  system measured. It will  vanish if the FCS is $x$-independent and a classical interpretation of the process is appropriate. 
 
We have seen that the FCS  is at least not directly observable in a single experiment. It is buried under some additional broadening which  we could fully understand in terms of the disturbance of the system by the detector. It therefore does not  tell us anything about fluctuations of the observable $A$ in our system, the quantity  we want to measure after all. In the next section we will therefore try  to get rid of this broadening and measure the function $P(x,q,T)$ itself, which is more specific to the system. A certain deconvolution procedure will be developed. 
\section{Characterization of the FCS; First scheme}
\label{sc:scheme1}
As we have already seen, the full counting statistics $P(x,q,T) $ proposed above allows us to predict the outcomes of measurements with a quite general detector and resolves inconsistencies that arose in earlier interpretations. It remains to show now, that it is a physical, real object in the sense that it is experimentally observable. 

We would like to suggest two schemes for measuring the FCS. We start from  relation (\ref{eq:dens}) between the initial and the final density matrix. Writing this equation in $(x^+,x^-)$-space, we find the simple expression
\begin{equation}
 P(x^+,x^-,T)= \frac{\rho^f_T(x^+,x^-)}{\rho^{in}(x^+,x^-)}
\end{equation}
or
\begin{equation}
 P(x,q,T) =  \int{ \frac{dz}{2 \pi} \; e^{iq z} \; \frac{\rho^f_T(x+z/2,x-z/2)}{\rho^{in}(x+z/2,x-z/2)}}
\end{equation}
for the kernel. We would already  be done if we could measure all elements of the detector's  final and  initial density matrices. This is not possible in general, however. By successively measuring a certain observable we can measure the diagonal elements of the density matrix in a  basis of eigenstates of that observable, but not the off-diagonal entries.  We can therefore measure the functions $\Pi(q)$, but not $\hat{\rho}$ itself. 

The key idea that we will pursue to solve this problem is to repeat the same measurement many times for shifted but otherwise identical initial detector density matrices. We suggest to repeat the measurement of the final momentum distribution $\Pi^f(q)$ for a number of initial density matrices that differ only in the expectation value $x_0$ of the position of the detecting particle and  call it
\begin{equation}
 \Gamma^f(x_0,q,T)\equiv \Pi^f_T(q,x_0) = \int{dx \; dq' \; P(x,q-q',T)\;\rho^{in}(x-x_0,q')} .
\end{equation}
This way we expose the  system during the measurement to different detector influences and one can hope that by doing so this  influence can be identified and eliminated by some kind of a deconvolution procedure. 
Defining the  Fourier transform of $ \Gamma^f(x_0,q,T)$ with respect to both of its variables
\begin{equation}
  \tilde{\Gamma}^f(q_0,z,T) \equiv \frac{1}{2\pi} \int{dx_0\;dq\; e^{ix_0q_0-iz q} \; \Gamma^f(x_0,q,T)}
\end{equation}
we find, that the FCS can indeed  be reconstructed from this function by means of the relation
\begin{equation} \label{eq:decon}
 P(x,q,T) =  \frac{1}{ 2 \pi} \int{dq_0\;dz\; e^{iqz - iq_0 x}\; \frac{ \tilde{\Gamma}^f(q_0,z,T)}{\tilde{\rho}^{in}(q_0,z)}} 
\end{equation}
where
 \begin{equation}
 \tilde{\rho}(q_0,z) \equiv \int{ dx \;e^{iq_0 x} \; \rho(x+\frac{z}{2},x-\frac{z}{2})} .
\end{equation} 
To interpret the result of the measurement, we  still have to know the full initial density matrix of the detector. This should be feasible, however. One might either prepare the detector initially in a specific, well-known state, or one might let the detector equilibrate with an environment. Then the initial density matrix is stationary, $0=[\hat{\rho}^{in},\hat{H}]\propto[\hat{\rho}^{in},\hat{q}^2]$, it will be diagonal in a basis of momentum eigenstates and can be determined by a momentum measurement only. 
This proves that the FCS really is  an observable.

To illustrate the procedure we want to apply it now to our example of the harmonic oscillator already introduced earlier.  Like there $\Pi^f(q)$ we calculate now the function $\Gamma^f(x_0,q,T)$ that would be measured in the actual experiment. We shift the initial oscillator wavefunction by $x_0$ in $x$-space and find that
\begin{equation}
 \Gamma^f(x_0,q,T) = \exp \bigl[\frac{-(q+x_0 T/M \omega^2)^2}{( 2 (\triangle q)^2 +  T^2/ 2 M^2 \omega^4 (\triangle q)^2 )} \bigr]. 
\end{equation}
On transforming this into Fourier space it becomes
\begin{equation}
\tilde{\Gamma}^f(q_0,z,T)  = \exp\bigl[-\frac{(\triangle q)^2 z^2 }{2} -\frac{T^2 z^2}{8 M^2 \omega^4 (\triangle q)^2}\bigr] \;\;\delta(q_0+  T  z/M \omega^2).
\end{equation}
Employing now Eq.\ (\ref{eq:decon}) with
$ \tilde{\rho^{in}}(q_0,z) =  \exp \bigl[-\frac{(\triangle q)^2 z^2}{2} -\frac{q_0^2}{8(\triangle q)^2} \bigr] $
 we indeed recover the desired FCS Eq.\ (\ref{eq:oscFCS}). 
\section{Second scheme}
\label{sc:scheme2}
We  see also a possibility of measuring $P(x,q,T)$ without knowledge of the initial state of the detector. We can apply it, however, only to the special case in which the system is in one state only, for example its ground state, or in a mixture of a limited number of discrete states. This condition may be satisfied   at low temperatures or if we have dissipation in the system which drives it into the ground state. We explain the idea for  the case of a system  in its ground state. Then we have the explicit expression (\ref{eq:ground}) for the time evolution and we find, that
\begin{equation}
 \Gamma^f(x_0,q,T) = \int{dx \; dz \;dq' \;  e^{-iz (q-q')  - iT (E'(x)\;z+\frac{1}{24} E'''(x)\;z^3 + ...) }\; \rho^{in}(x-x_0,q')} ,
\end{equation}
 $ \Gamma^f(x_0,q,T)$ again being the final momentum distribution for shifted initial detector wave functions.
In the limit of big $T$ we find with (\ref{eq:Pinfinite}) 
\begin{equation}
 \lim_{T \to \infty} \Gamma ^f(x_0,q,T) \propto \int{dx \;  \rho^{in}(x-x_0,q - TE'(x))}.
\end{equation}
This formula suggests that we can measure the function $E'(x)$ arbitrarily exact in the limit of long measurement time $T$ by determining the peak of the final momentum distribution. 
The only assumption we have to make about the initial detector density matrix now is, that it is well centered around $x=0$ and that it falls off sufficiently fast for momenta higher than some arbitrary $\triangle q$. We want $\hat{\rho}^{in}$ to be peaked in $x$-space such that $E'(x)$ is measured at the point $x_0$ only (we assume E(x) to be analytic). Of course, this means, that the width $\triangle q$ in momentum space of  $\hat{\rho}^{in}$ and therefore also of  $\hat{\rho}^{f}_T $ will be rather wide. This is not a problem, however, since the position of the peak  that we can measure with uncertainty $\triangle q$ increases linearly in time. The uncertainty in our knowledge of  $E'(x)$ is therefore determined  by$\frac{\triangle q}{T}$ and tends   to zero in the infinite time limit. 

Integrating $E'(x)$ we can then reconstruct the FCS for arbitrary detection times

$$ P(x,q,T)= \int{\frac{dz}{2 \pi}  \; \exp \{iqz  - iT  \int_{x-z/2}^{x+z/2}{dx' \; E'(x')} \}} $$ 

If the system is in a mixture of $N$ distinct states, the expression for  $P(x,q,T)$ will be a sum of terms of the form (\ref{eq:ground}) with different functions $E_n(x)$. We will in general find $N$ distinct peaks in the final momentum distribution allowing us to measure all $N$ functions $E_n'(x)$. Again, we can reconstruct  $P(x,q,T)$ for arbitrary $T$. 
\section{conclusions}
In this article we have been studying the process of measuring a general quantum mechanical variable $A$  by means of a detector that is coupled to it for a certain period of time. We have been focusing on a simple model describing a detector without internal dynamics which is nevertheless applicable to a variety of experimental situations. Moreover, the formalism that we have presented is general enough to allow for the description of a generic  detector. We have predicted the outcome of such a measurement. This prediction involves a new object that we want to call full counting statistics (FCS) of the variable $A$ and that is an extension of  another function proposed earlier in this context. 
This extension basically consists of accounting for the detector influence on the measured system. We find that the interplay of this influence with the quantum nature of the detector hampers in general a classical interpretation of the detector read-off. This way we have been able to  remove inconsistencies, to be more specific, negative probabilities, that arose in the previous interpretation. 
Finally, we have shown, that this FCS is not only a theoretical construct that helps to predict  results of certain measurements, but that it is an  observable itself. It can be measured in experiments. 
\acknowledgements
We have benefitted from discussions with   C.W.J. Beenakker, W. Belzig, and L.S. Levitov. This work was supported by the Dutch Science Foundation NWO/FOM.
\begin{appendix}
\section{}
In this appendix we give a detailed derivation of the infinite mass limit of the kernel (\ref{eq:kernel}).

First, we define a Fourier transformed influence functional
\begin{eqnarray}
 \tilde{F}[k^+(t),k^-(t),T] = \int{ D[x^+(t)] D[x^-(t)]  F[x^+(t),x^-(t),T]}\nonumber \\
 \exp\{{i} \int_{t_1}^{t_2}{dt\; \bigl[ x^+(t) k^+(t) - x^-(t) k^-(t)\bigr]} \} 
\end{eqnarray}
 
and correspondingly
\begin{equation}
 \tilde{K}(q^+,q^-,q_1^+,q_1^-,T) = \int{dx^+\;dx^- \; dx^+_1 \; dx^-_1 \exp\{ix^+q^+-ix^-q^--ix^+_1q^+_1 + ix^-_1q^-_1\} K(x^+,x^-,x_1^+,x_1^-,T)}
\end{equation}
The $k^{\pm}(t)$ are general functions on the interval $[t_1,t_2]$. 
Inserting the identity
\begin{equation}
 \exp \{{i} \int_{t_1}^{t_2}{ dt\; \frac{m}{2}\; {\dot{x}(t)}^2 }\}=\int{ D[q(t)]\; \exp\{{i} \int_{t_1}^{t_2}{dt\; \bigl[-\frac{q(t)^2}{2m} - q(t) \dot{x}(t)\bigr] } \}}
\end{equation}
and using (\ref{eq:kernel}) we derive then
\begin{eqnarray}
 \tilde{K}(q^+,q^-,q_1^+,q_1^-,T)=  \int{ \mathop{D[q^+(t)]}\limits_{q^+(t_2)=q^+,q^+(t_1)=q_1^+}\; \mathop{ D[q^-(t)]}\limits_{q^-(t_2)=q^-, q^-(t_1)=q_1^-}   } \nonumber \\
 \tilde{F}[\dot{q}^+(t),\dot{q}^-(t),T] \; \exp\{{i} \int_{t_1}^{t_2}{dt \;(\frac{{q^+(t)}^2}{2m}-\frac{{q^-(t)}^2}{2m})}\} 
\end{eqnarray}
In the infinite mass limit, the kinetic term in this expression will disappear. Now, we change integration variables from $D[q^{\pm}]$ to  $D[\dot{q}^{\pm}]$  and call $k^{\pm} = \dot{q}^{\pm}$. Then,

\begin{equation} \label{eq:p-kernel}
 \tilde{K}(q^+,q^-,q_1^+,q_1^-,T)=  \int{ \mathop{D[k^+(t)]}\limits_{\int_{t_1}^{t_2}{dt \;k^+(t)} = q^+-q_1^+} \; \mathop{D[k^-(t)]}\limits_{\int_{t_1}^{t_2}{dt \;k^-(t)} = q^--q_1^-}   \tilde{F}[k^+(t),k^-(t),T]} $$
\end{equation}

 We can represent the functions $k^{\pm}(t)$  by their Fourier series, $k^{\pm}(t)= \sum_{n=0}^{\infty}{ k_n^{\pm}\;\cos\;\frac{n \pi (t-t_1)}{t_2-t_1}}$. Changing the integration variables in (\ref{eq:p-kernel}) to the coefficients $k^{\pm}_n$ in this expansion, we notice, that only the integrals over the zeroth components $k^{\pm}_0$ are constrained by the boundary conditions.  We can therefore do all the integrals over higher Fourier modes in (\ref{eq:kernel}) and obtain  
\begin{eqnarray}
 \tilde{K}(q^+,q^-,q_1^+,q_1^-,T)= \int{ \mathop{D[k_n^+]}\limits_{n\neq0} \; \mathop{D[k_n^-]}\limits_{n\neq 0} \;D[x^+_n]\; D[x^-_n]} \nonumber \\
 F[x^+(t),x^-(t),T]\; \exp \{ \frac{i(t_2-t_1)}{2} \sum_{n=1}^{\infty} (k_n^+x_n^+ - k_n^-x_n^-) + {i} \bigl[ x_0^+(q^+-q_1^+)-x_0^-(q^--q_1^-)\bigr]\}
\end{eqnarray}
We have also expanded the functions $x^{\pm}(t)$ in Fourier series now,   $x^{\pm}(t)= \sum_{n=0}^{\infty}{ x_n^{\pm}\;\cos\;\frac{n \pi (t-t_1)}{t_2-t_1}}$.

We see, that the $k_n^{\pm}$-integrations result in  $\delta$-functions that constrain the $x_n^{\pm}$, $n \neq 0$, to zero and allow us to do the corresponding $x_n^{\pm}$ - integrals:
\begin{eqnarray}
 \tilde{K}(q^+,q^-,q_1^+,q_1^-,T)= \int{ dx_0^+ \; dx_0^- \;  \exp\{{i}\bigl[ x_0^+(q^+-q_1^+)-x_0^-(q^--q_1^-) \bigr] \}} \nonumber \\
 \mathop{Tr}\limits_{\rm System} T\; \exp\{-{i} \int_{t_1}^{t_2}{dt \bigl[ \hat{H}_{\rm sys}-\alpha_T(t)x^+_0 \hat{A}\bigr]}\}\;  \hat{R} \; \bar{T}\; \exp\{{i} \int_{t_1}^{t_2}{dt \bigl[\hat{H}_{\rm sys}-\alpha_T(t)x^-_0 \hat{A}\bigr]}\} 
\end{eqnarray}
or
\begin{equation}
 K(x^+,x^-,x_1^+,x_1^-,T)= \delta(x^+ - x_1^+)\; \delta(x^- - x_1^-) \; P(x^+,x^-,T) 
\end{equation}
with
\begin{equation}
 P(x^+,x^-,T) =  \mathop{Tr}\limits_{\rm System} T\; \exp\{-{i} \int_{t_1}^{t_2}{dt \bigl[\hat{H}_{\rm sys}-\alpha_T(t)x^+ \hat{A}\bigr]}\}\;  \hat{R} \; \bar{T}\;   \exp\{\;{i} \int_{t_1}^{t_2}{dt \bigl[ \hat{H}_{\rm sys}-\alpha_T(t)x^-\hat{A}\bigr]}\} .
\end{equation}
Thus, the claimed locality of  the kernel $ K(x^+,x^-,x_1^+,x_1^-,T)$ is established.
 
\end{appendix}

\end{document}